\pdfoutput=1

\documentclass[a4paper,cits]{JINST}
\usepackage{microtype}
\usepackage[hyphens]{url}

\title{In-situ calibration of a PMT inside a scintillation detector by means of primary scintillation detection}

\author{
The NEXT Collaboration

V.~\'Alvarez,$^{a}$
F.I.G.~Borges,$^{b}$
S.~C\'arcel,$^{a}$
J.~Castel,$^{c}$
S.~Cebri\'an,$^{c}$
A.~Cervera,$^{a}$
C.A.N.~Conde,$^{b}$
T.~Dafni,$^{c}$
T.H.V.T.~Dias,$^{b}$
J.~D\'iaz,$^{a}$
M.~Egorov,$^{d}$
R.~Esteve,$^{e}$
P.~Evtoukhovitch,$^{f}$
L.M.P.~Fernandes,$^{b}$
P.~Ferrario,$^{a}$
A.L.~Ferreira,$^{g}$
E.D.C.~Freitas,$^{b}$\thanks{Corresponding author}~
V.M.~Gehman,$^{d}$
A.~Gil,$^{a}$
A.~Goldschmidt,$^{d}$
H.~G\'omez,$^{c}$
J.J.~G\'omez-Cadenas,$^{a}$
D. Gonz\'alez-D\'iaz,$^{c}$
R.M.~Guti\'errez,$^{h}$
J.~Hauptman,$^{i}$
J.A.~Hernando Morata,$^{j}$
D.C.~Herrera,$^{c}$
F.J.~Iguaz,$^{c}$
I.G.~Irastorza,$^{c}$
M.A.~Jinete,$^{h}$
L.~Labarga,$^{k}$
A.~Laing,$^{a}$
I.~Liubarsky,$^{a}$
J.A.M.~Lopes,$^{b}$
D.~Lorca,$^{a}$
M.~Losada,$^{h}$
G.~Luz\'on,$^{c}$
A.~Mar\'i,$^{e}$
J.~Mart\'in-Albo,$^{a}$
A.~Mart\'inez,$^{a}$
T.~Miller,$^{d}$
A.~Moiseenko,$^{f}$
F.~Monrabal,$^{a}$
C.M.B.~Monteiro,$^{b}$
F.J.~Mora,$^{e}$
L.M. Moutinho,$^{g}$
J.~Mu\~noz~Vidal,$^{a}$
H.~Natal da Luz,$^{b}$
G.~Navarro,$^{h}$
M.~Nebot,$^{a}$
D.~Nygren,$^{d}$
C.A.B.~Oliveira,$^{dg}$
R.~Palma,$^{l}$
J.~P\'erez,$^{k}$
J.L.~P\'erez~Aparicio,$^{l}$
J.~Renner,$^{d}$
L.~Ripoll,$^{m}$
A.~Rodr\'iguez,$^{c}$
J.~Rodr\'iguez,$^{a}$
F.P.~Santos,$^{b}$
J.M.F.~dos~Santos,$^{b}$
L.~Segu\'i,$^{c}$
L.~Serra,$^{a}$
D.~Shuman,$^{d}$
A.~Sim\'on,$^{a}$
C.~Sofka,$^{n}$
M.~Sorel,$^{a}$
J.F.~Toledo,$^{d}$
A.~Tom\'as,$^{c}$
J.~Torrent,$^{m}$
Z.~Tsamalaidze,$^{f}$
D.~V\'azquez,$^{j}$
J.F.C.A.~Veloso,$^{g}$
J.A.~Villar,$^{c}$
R.~Webb,$^{n}$
J.T.~White,$^{n}$
N.~Yahlali$^{a}$\\
\llap{$^{a}$}
Instituto de F\'isica Corpuscular (IFIC), CSIC \& Universitat de Val\`encia\\
Calle Catedr\'atico Jos\'e Beltr\'an, 2, 46980 Paterna, Valencia, Spain\\
\llap{$^{b}$}
Departamento de Fisica, Universidade de Coimbra\\
Rua Larga, 3004-516 Coimbra, Portugal\\
\llap{$^c$}
Lab.\ de F\'isica Nuclear y Astropart\'iculas, Universidad de Zaragoza\\ 
Calle Pedro Cerbuna, 12, 50009 Zaragoza, Spain\\
\llap{$^d$}
Lawrence Berkeley National Laboratory (LBNL)\\
1 Cyclotron Road, Berkeley, California 94720, USA\\
\llap{$^{e}$}
Instituto de Instrumentaci\'on para Imagen Molecular (I3M), Universitat Polit\`ecnica de Val\`encia\\ 
Camino de Vera, s/n, Edificio 8B, 46022 Valencia, Spain\\
\llap{$^{f}$}
Joint Institute for Nuclear Research (JINR)\\
Joliot-Curie 6, 141980 Dubna, Russia\\
\llap{$^{g}$}Institute of Nanostructures, Nanomodelling and Nanofabrication (i3N), Universidade de Aveiro\\
Campus de Santiago, 3810-193 Aveiro, Portugal\\
\llap{$^{h}$}
Centro de Investigaciones, Universidad Antonio Nari\~no\\ 
Carretera 3 este No.\ 47A-15, Bogot\'a, Colombia\\
\llap{$^{i}$}
Department of Physics and Astronomy, Iowa State University\\
12 Physics Hall, Ames, Iowa 50011-3160, USA\\
\llap{$^{j}$}
Instituto Gallego de F\'isica de Altas Energ\'ias (IGFAE), Univ.\ de Santiago de Compostela\\
Campus sur, R\'ua Xos\'e Mar\'ia Su\'arez N\'u\~nez, s/n, 15782 Santiago de Compostela, Spain\\
\llap{$^{k}$}
Departamento de F\'isica Te\'orica, Universidad Aut\'onoma de Madrid\\
Ciudad Universitaria de Cantoblanco, 28049 Madrid, Spain\\
\llap{$^{l}$}
Dpto.\ de Mec\'anica de Medios Continuos y Teor\'ia de Estructuras, Univ.\ Polit\`ecnica de Val\`encia\\
Camino de Vera, s/n, 46071 Valencia, Spain\\
\llap{$^{m}$}
Escola Polit\`ecnica Superior, Universitat de Girona\\
Av.~Montilivi, s/n, 17071 Girona, Spain\\
\llap{$^{n}$}
Department of Physics and Astronomy, Texas A\&M University\\
College Station, Texas 77843-4242, USA\\

E-mail: \email{elisabete@gian.fis.uc.pt}
}

\abstract{
We have investigated the possibility of calibrating the PMTs of scintillation detectors using the primary scintillation produced by X-rays to induce single photoelectron response of the PMT. The high-energy tail of this response can be approximated to an exponential function under some conditions. In these cases, it is possible to determine the average gain for each PMT biasing voltage from the inverse of the exponent of the exponential fit to the tail, which can be done even if the background and/or noise cover-up most of the distribution. We have compared our results with those obtained by the commonly used single electron response (SER) method, which uses a LED to induce a single photoelectron response of the PMT and determines the peak position of such response, relative to the pedestal peak (the electronic noise peak, which corresponds to 0 photoelectrons). The results of the exponential fit method agree with those obtained by the SER method when the average number of photoelectrons reaching the first dynode per light/scintillation pulse is around 1.0. The SER method has higher precision, while the exponential fit method has the advantage of being useful in situations where the PMT is already in situ, being difficult or even impossible to apply the SER method, e.g. in sealed scintillator/PMT devices.
}

\preprint{\today} 

\begin{document}

\tableofcontents

\section{Introduction} \label{sec:Introduction}
The knowledge of the PMT gain can be important in many different situations. Up to the present, PMTs have been widely used as photosensors of scintillation detectors, which are extensively applied to $\gamma$-ray spectrometry, particle detection for high-energy physics and rare event detection. The knowledge of the PMT gain allows the determination of the number of photoelectrons released by the photocathode, following the measurement of the total number of electrons collected in the anode. The determination of the number of photoelectrons is an important parameter for the absolute studies of the scintillator response to electromagnetic and charged particles and the respective dependence on energy. In particular, the absolute energy dissipated in the scintillators, is related to the number of the photons produced, which in turn is related to the number of photoelectrons emitted by the PMT photocathode. 

An effective method for the PMT gain determination is achieved obtaining the PMT response to single photoelectron (SER) \cite{Bellamy:1994bv, Dossi:1998zn, ChirikovZorin:2001qv}: Usually, a LED is used to irradiate the PMT and the amount of light reaching the PMT is reduced in order to induce a single photoelectron response of the PMT. The position of the peak produced by single photoelectron emission is measured relative to the position of the pedestal peak (the electronic noise peak, which corresponds to zero induced photoelectrons). If the amount of light is small enough the single photoelectron emission is dominant over the multiple photoelectron emission cases. The peak distribution associated to events resulting from a given number of photoelectrons reaching the first dynode can be approximated to a Gaussian and the relative position of the centroid of this Gaussian presents a linear increase with the number of photoelectrons hitting the first dynode, while the value of the Gaussian area, which is related with the probability for the corresponding event, obeys to a Poisson distribution \cite{Bellamy:1994bv, Dossi:1998zn, ChirikovZorin:2001qv}. Therefore, either the SER is deconvoluted in order to determine the position of the peak corresponding to events with a single photoelectron emission, or the amount of the LED light hitting the PMT is reduced to a point where the probability of having multiple photoelectron emission is less than few percent, when compared to the probability for the single photoelectron emission, being the peak well described by a single Gaussian function.

In many experiments, the PMT Gain is measured before assembling (e.g. see \cite{Bueno:2007au}) or even by placing one or more LEDs and/or optical fibres inside the detector, in order to allow the monitoring of the PMT gain with time, along the experiment (e.g. see \cite{Aprile:2011dd}). However, there are many cases where the PMT is inside a sealed chamber inaccessible to a LED light, such is the case of most scintillation detectors, being impossible to use the above method.
In this work, we have investigated the possibility of calibrating a PMT of a scintillation detector using the primary scintillation produced by the radiation interaction in the scintillator. If the primary scintillation induces a single photoelectron response of the PMT and an exponential function may be fitted to the high-energy tail of the PMT pulse-height distribution, the PMT average gain can be determined from the inverse of the exponent of the exponential fit. In fact, observations have already been made under some conditions, which show exponential-like distributions \cite{Knoll, Prescott:1966, Inman:1969}.

\section{Experimental setup and method} \label{sec:Setup}

\begin{figure}
\centering
\includegraphics[scale=1]{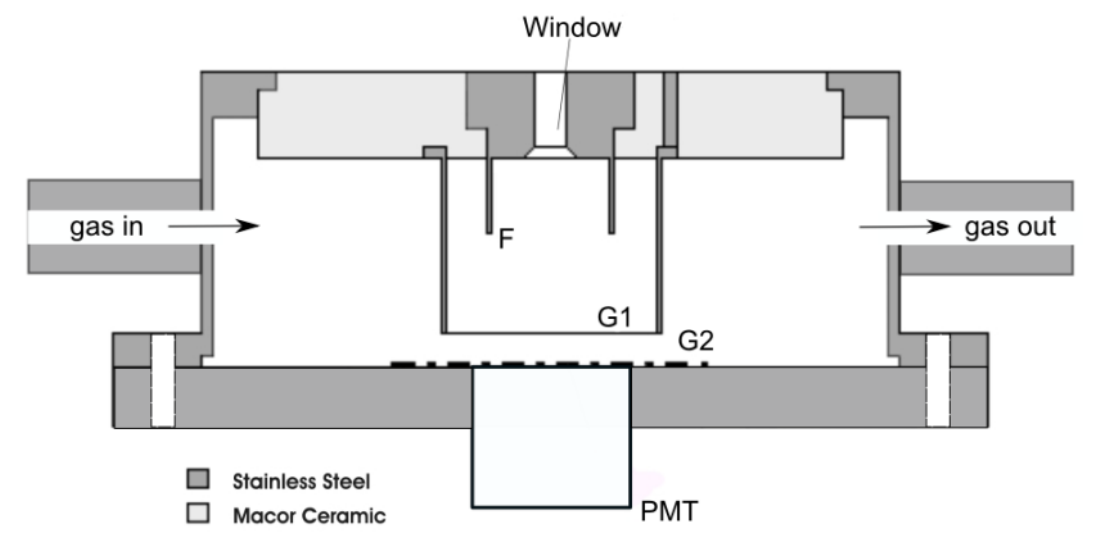}
\caption{Schematic of the xenon GPSC with a R8520-06SEL PMT used as VUV scintillation photosensor.}\label{fig:Setup}
\end{figure}

The Gas Proportional Scintillation Counter (GPSC) used in this work was built to study the performance of a Hamamatsu R8520-06SEL photomultiplier for readout primary and secondary (electroluminescence) scintillation produced as a result of X-ray absorptions in xenon, at room temperature \cite{Fernandes:2010gg}. The GPSC schematic is presented in figure \ref{fig:Setup} and it was described in detail in \cite{Fernandes:2010gg}. 

The GPSC body is made of stainless steel. The R8520-06SEL PMT was glued with low vapour pressure epoxy (TRA-CON 2116) to the anode plane, the bottom base of the GPSC body, fig.~\ref{fig:Setup}. The GPSC upper base is a Macor disk having an aluminized Kapton X-ray window insulated from the detector body. The Macor is epoxied to the X-ray window holder and to the GPSC body using the epoxy already refereed to above. The bottom base is vacuum sealed to the GPSC upper part by compressing an indium gasket, allowing detector disassembling when necessary. The drift/conversion region is 3 cm thick and it is located between the window and mesh G1, while the electroluminescence gap, 0.5 cm thick, is located between mesh G1 and the anode plane (mesh G2), placed on top of the PMT quartz window. 

A $^{55}$Fe radioactive source is positioned outside the chamber, on top of the detector window and a 1mm diameter lead collimator. A thin film of chromium was placed between the radioactive source and the window to efficiently reduce the interactions of 6.4 keV X-rays (Mn K$\beta$ line) in the gas volume. Primary scintillation resulting from the 5.9 keV X-ray interactions is produced in the xenon during the formation of the primary electron cloud, following the absorption of the X-ray and the subsequent thermalisation of the photoelectron and other Auger electrons emitted by the atom. The amplitude of primary scintillation pulses is very low, difficult to distinguish from noise or background pulses.

\begin{figure}
\centering
\includegraphics[width=0.49\textwidth]{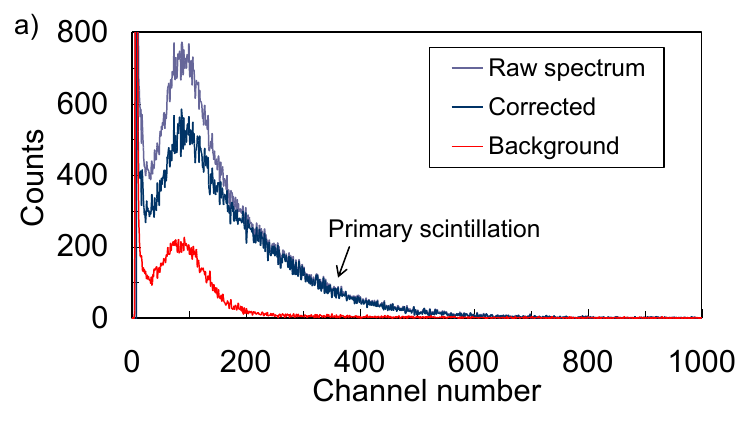}
\includegraphics[width=0.49\textwidth]{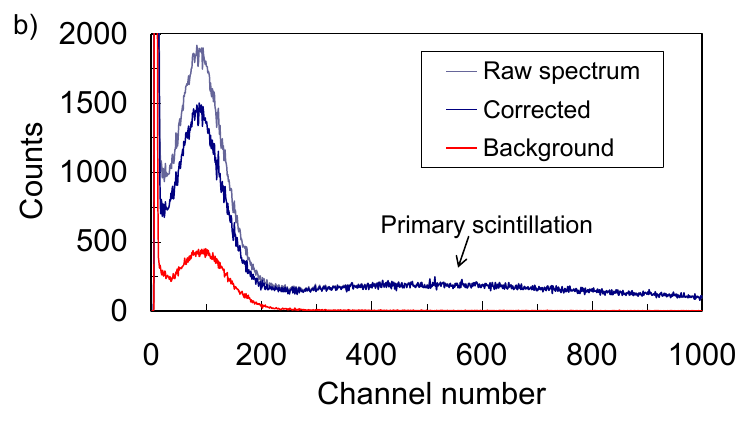}
\caption{Pulse-height distribution obtained in the detector, with no electric fields applied to the drift and scintillation regions and for a PMT bias voltage of 730 V: a) for 5.9 keV X-rays and b) for the X-rays emitted from a $^{109}$Cd radioactive source. }\label{fig:Fig2}
\end{figure}

Figure \ref{fig:Fig2}a depicts the pulse-height distribution recorded in the MCA for the primary scintillation resulting from 5.9 keV X-rays interactions in xenon, with no electric fields applied to the GPSC drift and scintillation regions. In order to access the contribution of background pulses resulting from the PMT dark current and/or from residual visible light entering the chamber, energy spectra with and without X-ray irradiation were recorded. This way, the pulse-height distribution for that background could be identified and subtracted from the raw spectrum. The count rate of this background decreases by improving the light shielding of the detector. In figure \ref{fig:Fig2}b, we present the pulse-height distribution obtained irradiating the detector with 22.1 and 25.0 keV X-rays from a $^{109}$Cd radioactive source. As shown, a similar distribution is obtained in the low-energy region, while a broader pulse-height distribution extends towards the high-energy region. This indicates that the pulse-height distribution obtained after the referred to above background subtraction has other contributions than the one resulting from the detection of the primary scintillation. We believe that the peak obtained in the low energy region results from luminescence and/or fluorescence of the detector materials, including the PMT, induced by the presence of the X-ray and/or VUV photon interactions, being the distribution due to primary scintillation superimposed to this peak, extending towards the higher energy region.

\begin{figure}
\centering
\includegraphics[scale=1]{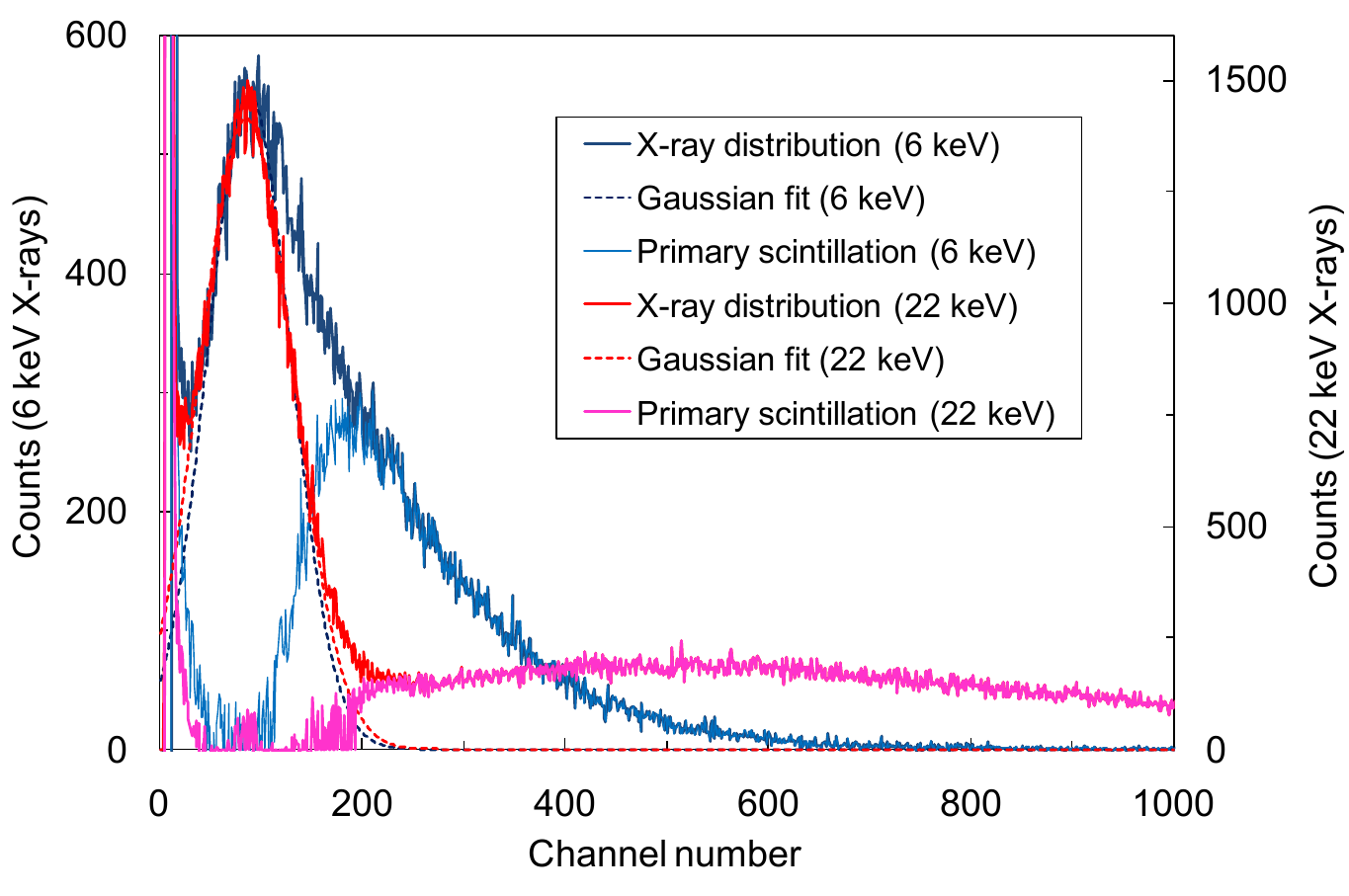}
\caption{Estimated pulse height distribution of primary scintillation pulses obtained by subtraction of a Gaussian curve fitted to the low-energy region of the distribution. The final distribution is shown for the primary scintillation resulting from 5.9 keV and for $^{109}$Cd keV X-ray absorptions in xenon.}\label{fig:Fig3}
\end{figure}

Assuming a Gaussian shape for the low energy peak, we can subtract it from the pulse-height distribution in order to obtain the part of the primary scintillation pulse-height distribution that is outside of this background peak, figure \ref{fig:Fig3}. In a former study \cite{Fernandes:2010gg}, we have shown that this component and, therefore, the primary scintillation is independent of the electric field present on the conversion region and independent of the xenon pressure. In addition, we have measured the Xe primary scintillation yield and the average number of primary scintillation photons hitting the PMT active area per 5.9 keV X-ray absorptions. In average, 82 Xe VUV photons are produced in per 5.9 keV X-ray absorption in xenon and only an average of 2 of them reach the PMT active area, in our setup. Since the quantum efficiency of the present PMT, which has a QE enhanced photocathode, is about 33\% for the Xe VUV scintillation \cite{Hamamatsu}, the 5.9 keV X-ray interactions in our detector leads to a pulse-height distribution resulting from a single photoelectron response of the PMT.

\section{Experimental results and discussion}
As referred to above, part the pulse-height distribution resulting from the PMT response to single photoelectron detection is under the background low energy peak and only the tail of this distribution is out of this peak, as shown in fig.~\ref{fig:Fig3}. Therefore, in this case, it is not possible to determine the PMT gain using the SER method described above. Nevertheless, if we can approximate the single photoelectron response of the PMT to an exponential function, 
\begin{equation}
P(q) = a\ \exp{\left(-q/q_\mathrm{avg}\right)}
\end{equation}
being $q$ the charge collected at the PMT anode and $q_\mathrm{avg}$ the average charge obtained from all the events, only the high-energy tail of the distribution is, then, necessary to fit in order to extract the PMT average gain. We can determine the exponent of this distribution, which presents a linear trend with q when represented in log scale, being the slope of this tail, i.e. the exponent, the inverse of the PMT average gain. This approximation can be possible under some conditions \cite{Knoll, Prescott:1966, Inman:1969}.

\begin{figure}
\centering
\includegraphics[scale=1]{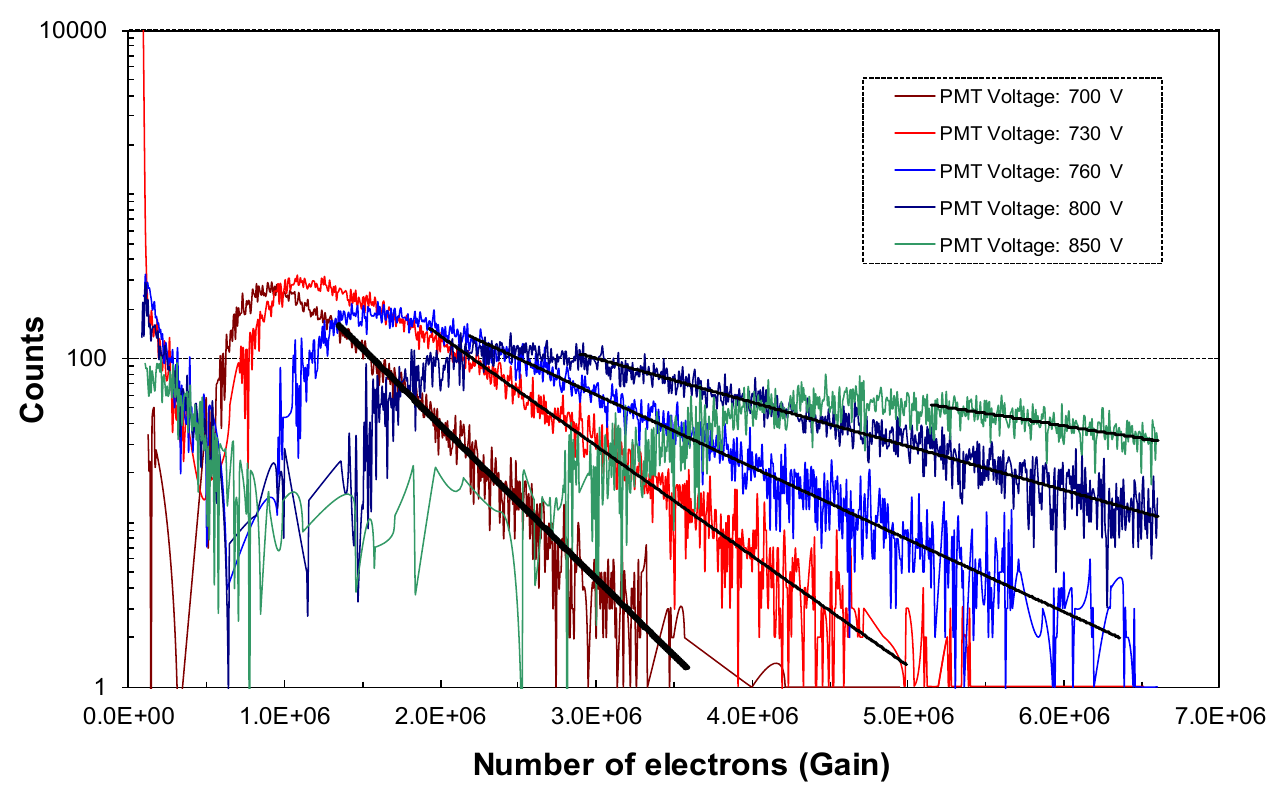}
\caption{Pulse-height distributions resulting from 5.9 keV X-ray interactions in the detector, obtained after the subtraction of the low energy background peak, for different PMT biasing voltages.}\label{fig:Fig4}
\end{figure}

Figure \ref{fig:Fig4} depicts the different pulse-height distributions of the primary scintillation resulting from 5.9 keV X-ray interactions in Xe, for different PMT biasing voltages. These distributions were obtained after the subtraction of the low energy background peak referred to above. The solid lines represent the exponential fits to the tails of the obtained pulse-height distributions, in the high-energy region. All the tails are very well described by an exponential function and, as the PMT voltage increases, the distributions extend to higher amplitudes with a smaller slope. The MCA channel number was calibrated in number of electrons using a pulse generator to feed a calibrated capacitor directly coupled to the pre-amplifier input. 

\begin{figure}
\centering
\includegraphics[scale=1]{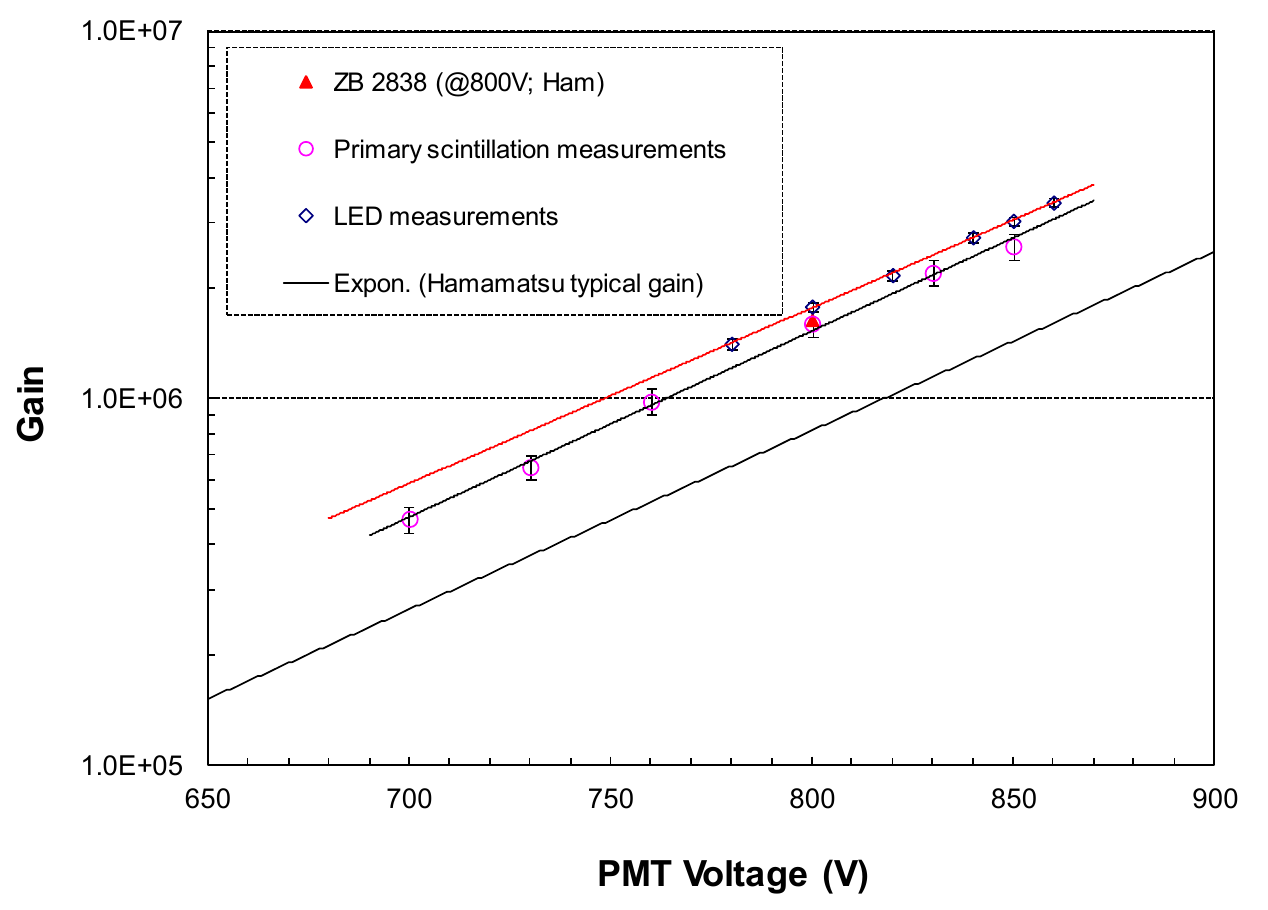}
\caption{PMT Gain obtained using the SER method (blue diamonds) and using the exponential fit to the pulse-height distributions resulting from the detection of primary scintillation (pink circles). Also depicted is the value obtained by Hamamatsu at 800V (red triangle) and the typical gain for the standard, no QE enhanced, R8520 PMT, according to Hamamatsu.}\label{fig:Fig5}
\end{figure}

The experimental results obtained for the PMT gain are depicted in figure \ref{fig:Fig5}. To confirm the obtained results, we have disassembled the detector and measured the PMT gain using the SER technique. These latter values are also depicted in figure \ref{fig:Fig5} for comparison, together with the value measured by Hamamatsu at 800V and supplied in the PMT data-sheet, as well as the typical gain for the standard R8520 PMT series (not having a QE enhanced photocathode). In our setup for the studies with the LED, the SER method can not be applied for PMT voltages below 780 V, because the single-photoelectron peak of the pulse-height distribution gets under the electronic noise (pedestal) peak, while for the exponential tail measurements we can measure gains for PMT voltages lower than 700V. The results obtained using the exponential tail agree with those obtained using the SER method within 15\%. However, a systematic deviation towards lower gain values is observed. The value presented by Hamamatsu is for the effective PMT gain, i.e. including the collection efficiency of photoelectrons by the first dynode, an effect that is not included in the other values since they measure the effective charge gain through the dynode chain.

\begin{figure}
\centering
\includegraphics[scale=1]{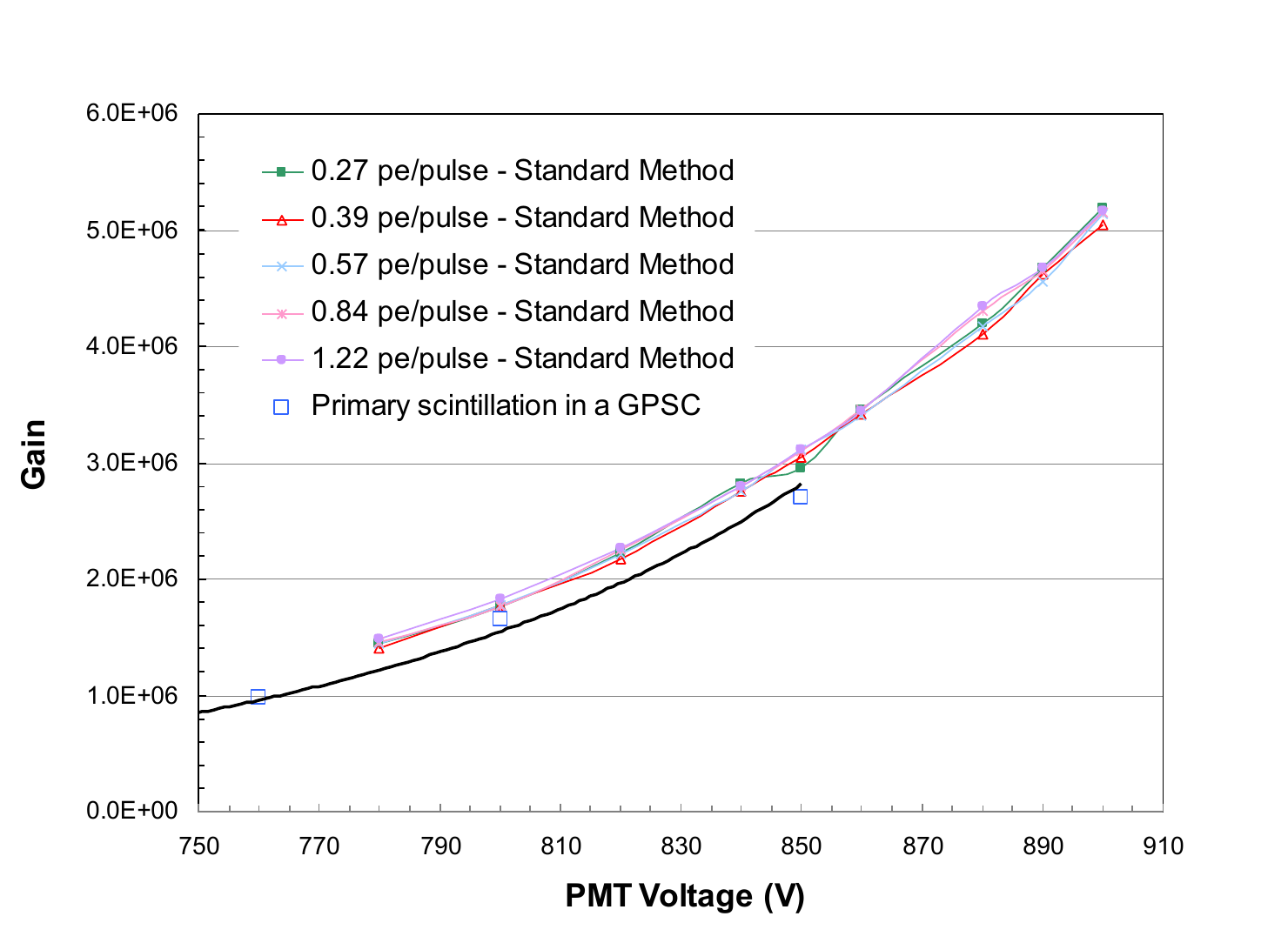}
\caption{PMT gain obtained as a function of the PMT biasing voltage, using the SER method and for different LED light levels, corresponding to different average number of photoelectrons emitted by the photocathode. For comparison, the PMT gain obtained by the exponential fit method, using the Xe primary scintillation, is also presented.}\label{fig:Fig6}
\end{figure}

This systematic difference could be due to the different amount of light hitting the PMT in primary scintillation and in LED light measurements. While in primary scintillation measurements the amount of light hitting the PMT active area corresponds to a photocathode average emission of $\sim0.7$ photoelectrons per scintillation pulse, this number is usually set below 0.1 for the LED light irradiation technique, in order to reduce the contribution of multiple photoelectron emission by the PMT photocathode to a negligible level. Therefore, to understand the effect of the amount of light detected by the PMT on the measurements using the SER method, i.e. on the determination of centroid of the single photoelectron peak, we have measured the PMT gain versus the PMT biasing voltage for different levels of light emitted by the LED, in order to change the corresponding average photoelectron emission per light pulse between 0.1 and 1.2. The results are presented in figure \ref{fig:Fig6}. As shown, the gains obtained by the SER method do not depend on the LED light level, within the studied range. For comparison, the PMT gain obtained by the exponential fit method, using the Xe primary scintillation, is also presented.

\begin{figure}
\centering
\includegraphics[scale=1]{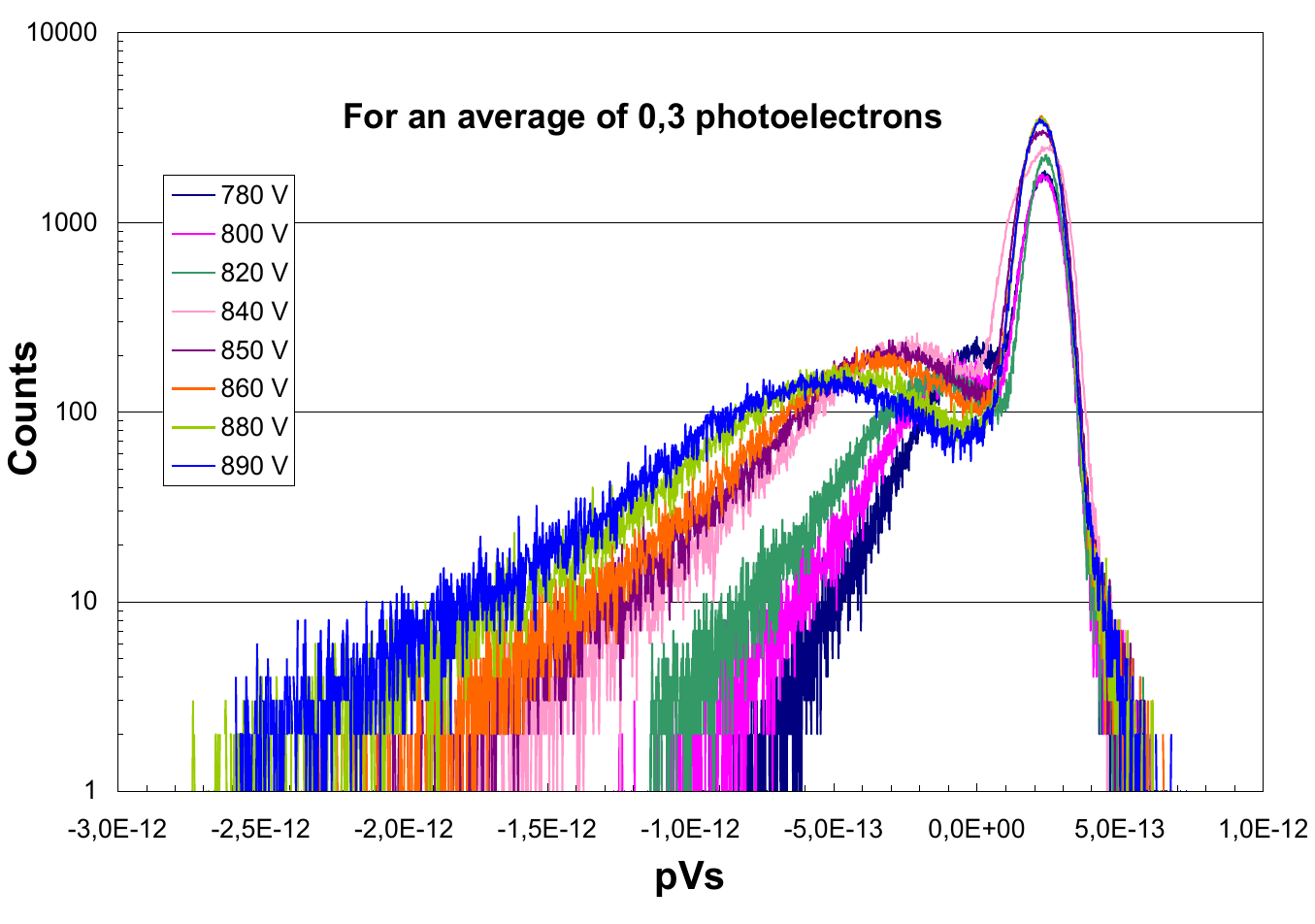}
\includegraphics[scale=1]{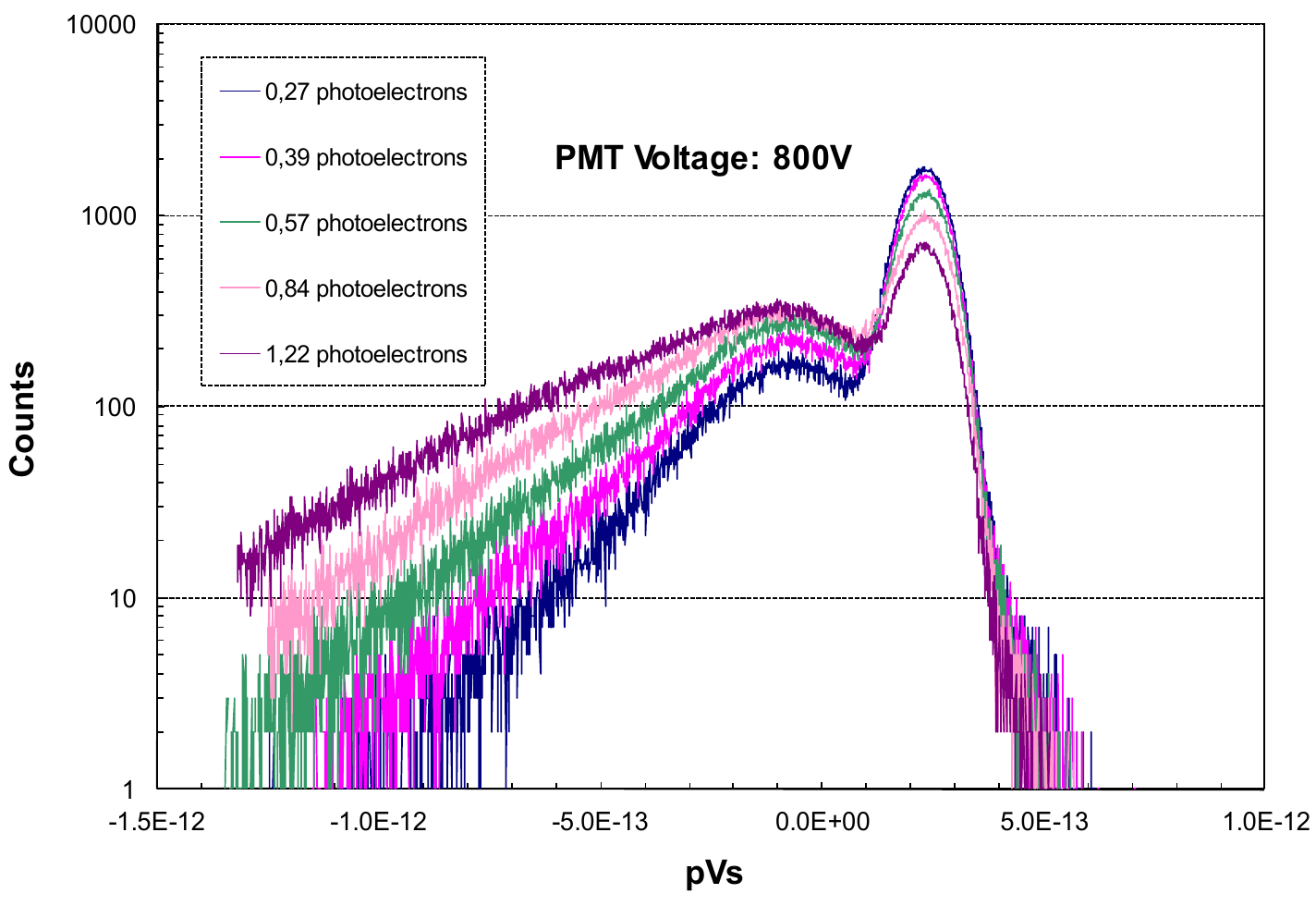}
\caption{PMT pulse height distributions: (top) for a constant LED illumination level corresponding to an average of 0.3 photoelectrons emitted by the photocathode per light pulse and different PMT biasing voltages; (bottom) for a constant PMT biasing voltage of 800 V and different LED illumination levels.}\label{fig:Fig7}
\end{figure}

After this study, we have turned to the exponential fit method and we have applied it to the pulse-height distributions obtained with the LED irradiation for different PMT voltages and for different LED light levels. Figure~\ref{fig:Fig7}~(top) depicts typical pulse height distributions obtained for different PMT biasing voltages and for a LED illumination level corresponding to an average of 0.3 photoelectrons emitted by the photocathode per light pulse, while fig.~\ref{fig:Fig7}~(bottom) depicts typical pulse height distributions obtained for different LED illumination levels and a constant PMT biasing voltage of 800 V.

\begin{figure}
\centering
\includegraphics[scale=1]{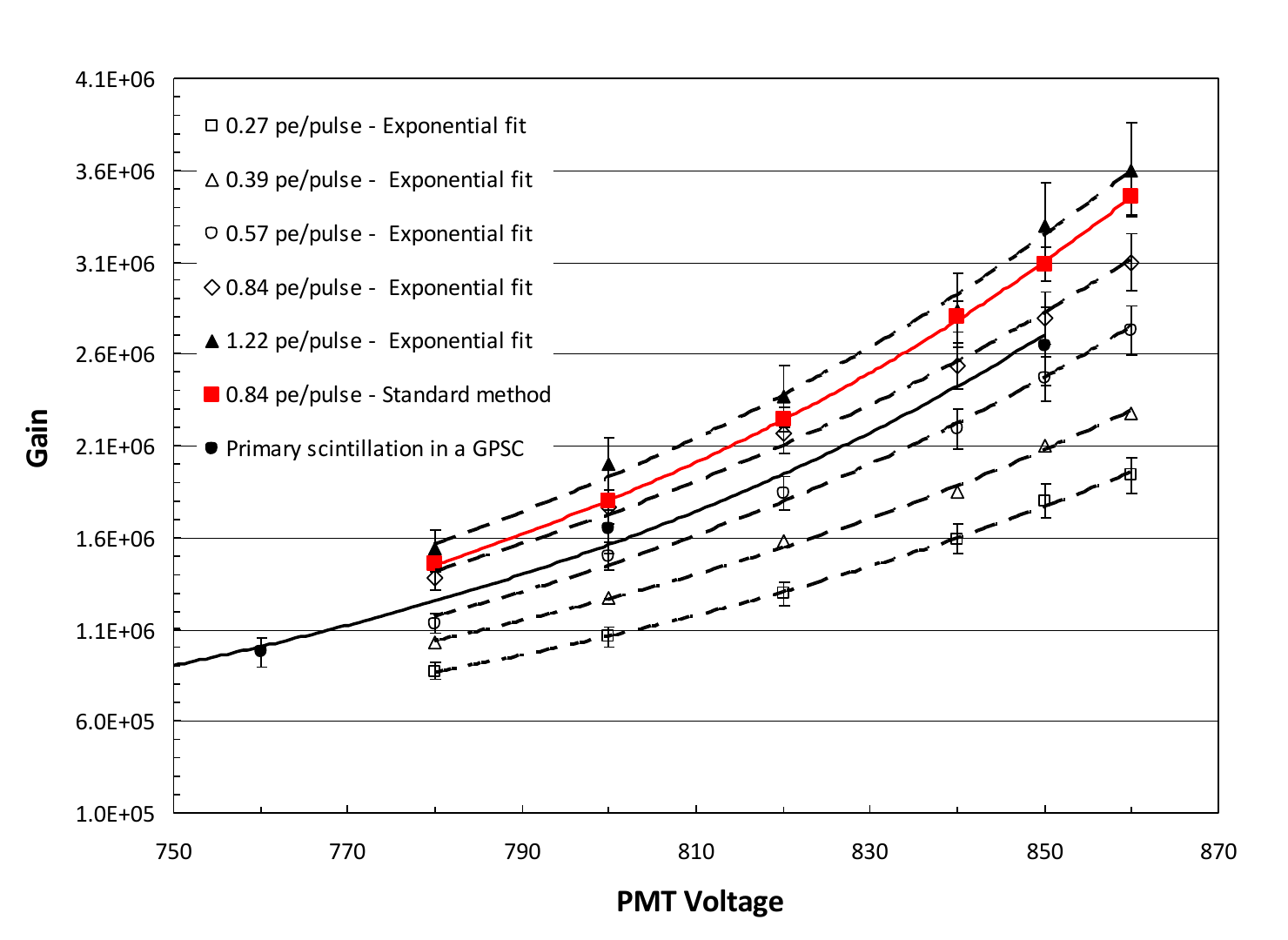}
\caption{PMT average gain as a function of the PMT biasing voltage, obtained by the exponential fit to the tails of the SER pulse-height distributions resulting from the LED illumination with different light levels, corresponding to different average number of photoelectrons emitted by the photocathode. For comparison, the PMT average gain obtained by the exponential fit to the pulse-height distributions from Xe primary scintillation, as well as the PMT gain obtained through the SER single photoelectron peak determination, are also presented. }\label{fig:Fig8}
\end{figure}

Figure \ref{fig:Fig8} depicts the PMT average gain as a function of the PMT bias voltage for different LED light illumination levels, obtained through the exponential fit to the high-energy tails of the respective SER pulse-height distributions. For comparison, The PMT average gain obtained using the SER method as well as the average gain obtained using the exponential fit to the pulse-height distributions of the Xe primary scintillation induced by 5.9 keV X-Rays are also depicted. As seen, the PMT gain obtained by the exponential fit method depends on the LED light level, i.e. on the amount light hitting the photocathode. Nevertheless, relative to the gain obtained measuring the single photoelectron peak, the obtained gains are within 10\% for LED light levels in a range corresponding to an average of 0.8--1.2 photoelectrons amplified in the PMT dynode chain per light pulse.

This is not surprising taking into account that for low light illumination, e.g. below an average of 0.1 photoelectrons collected in the first dynode per light pulse, the SER pulse-height distribution of the PMT is well described by a Gaussian function \cite{Bellamy:1994bv, Dossi:1998zn, ChirikovZorin:2001qv}. As the amount of light illuminating the PMT increases, the probability for multiple photoelectron emission increases and the pulse-height distribution consists of the sum of the different Gaussians corresponding to events with different number of photoelectrons hitting the first dynode. The centroid of each Gaussian presents a linear increase with the number of photoelectrons hitting the first dynode, while the value of the Gaussian area, which is related with the probability for the corresponding event, obeys to a Poisson distribution \cite{Bellamy:1994bv, Dossi:1998zn}. Therefore, the sum of this multiple Gaussians results in a pulse-height distribution having a high-energy tail that can be approximated to an exponential function. Figure~\ref{fig:Fig8} shows that if the light pulse induces an average photoelectron emission around 1.0 the tail of the pulse height distribution can be approximated to an exponential function and the PMT average gain can be obtained by the inverse of the exponent of this function. Lower scintillation levels will lead to lower gain values, when comparing to those obtained with the SER method. Nevertheless, the measured gain is still within a factor of two for scintillation levels inducing, in average, as low as 0.3 photoelectrons per scintillation pulse.

The proposed method has been applied successfully by the NEXT Collaboration to monitor the gain variations of the PMTs used by the NEXT-DEMO detector \cite{Alvarez:2012nd}. The reference gains in NEXT-DEMO are obtained using the SER method, but the evolution of the gains with time is followed using the method described in this paper.

\section{Conclusions}
We have measured the average gain of a PMT assembled inside a xenon Gas Proportional Scintillation Counter by registering the pulse height distributions resulting from the primary scintillation light produced in xenon due to 5.9 keV X-Ray interactions. As these primary scintillation pulses induce, in average, about 0.7 photoelectrons collected in the PMT dynode chain, we have assumed that the charge avalanche in the PMT could be described by an exponential function, being the PMT average gain given by the inverse of the exponent. The results obtained are consistent but about 15\% lower than those obtained using the SER method, i.e. using LED light pulses to induce a single photoelectron response in the PMT and determining the peak of such response relative to the centroid of the noise peak (pedestal).

The present studies have shown that PMT gain calibration obtained through the fit of an exponential function to the high-energy tail of the PMT SER pulse-height distributions is a valid procedure, for PMT illuminations within suitable levels, and best results are obtained if the primary scintillation produced by the radiation induces in average around 1.0 photoelectrons per scintillation pulse. Lower scintillation levels will lead to lower gain values, when comparing to those obtained with the SER method. Nevertheless, the measured gain is still within a factor of two for scintillation levels inducing in average as low as 0.3 photoelectrons per scintillation pulse.

This method is very useful when the PMT is assemble and sealed together with the scintillator, like in many scintillation counters, and cannot be removed, being impossible to use the SER method. Also, this method can be used even when the detector background noise is high, covering the single electron response peak, as it was in the present experimental case, since only the high-energy tail is needed.

\acknowledgments
This work was supported by the Portuguese FCT and FEDER through the program COMPETE, projects PTDC/FIS/103860/2008, and by the Spanish Ministerio de Econom\'ia y Competitividad under grants CONSOLIDER-Ingenio 2010 CSD2008-0037 (CUP) and FPA2009-13697-C04-04.

\bibliographystyle{JHEP}
\bibliography{references}

\end{document}